\documentclass[conference]{IEEEtran}

\usepackage{fancyhdr}

\usepackage{soul}  %% underline wrap long line
\usepackage[detect-all]{siunitx}
\usepackage{array} %% tab align
\usepackage{booktabs} %% \toprule 
\usepackage{multirow}
\usepackage[export]{adjustbox}

\newcommand{\figscale}{0.95}

\IEEEoverridecommandlockouts
\usepackage{cite}
\usepackage{amsmath,amssymb,amsfonts}
\usepackage{algorithmic}
\usepackage{graphicx}
\usepackage{textcomp}
\usepackage{xcolor}
\usepackage{tcolorbox}
\def\BibTeX{{\rm B\kern-.05em{\sc i\kern-.025em b}\kern-.08em
    T\kern-.1667em\lower.7ex\hbox{E}\kern-.125emX}}
\begin{document}

\title{
Study of Workload Interference with Intelligent Routing on Dragonfly
}

\author{
\IEEEauthorblockN{Yao Kang}
\IEEEauthorblockA{\textit{Department of Computer Science} \\
\textit{Illinois Institute of Technology}\\
Chicago, USA \\
ykang17@hawk.iit.edu}
\and
\IEEEauthorblockN{Xin Wang}
\IEEEauthorblockA{\textit{Department of Computer Science} \\
\textit{Illinois Institute of Technology}\\
Chicago, USA \\
xwang149@hawk.iit.edu}
\and
\IEEEauthorblockN{Zhiling Lan}\thanks{Zhiling Lan's current affiliation is University of Illinois Chicago, and her current contact is \ttfamily\upshape{zlan@uic.edu}.}
\IEEEauthorblockA{\textit{Department of Computer Science} \\
\textit{Illinois Institute of Technology}\\
Chicago, USA \\
lan@iit.edu}
}

\maketitle

\thispagestyle{fancy}
\lhead{}
\rhead{}
\chead{}
\lfoot{\footnotesize{
SC22, November 13-18, 2022, Dallas, Texas, USA
\newline 978-1-6654-5444-5/22/\$31.00 \copyright 2022 IEEE}}
\rfoot{}
\cfoot{}
\renewcommand{\headrulewidth}{0pt}
\renewcommand{\footrulewidth}{0pt}

\begin{abstract}
Dragonfly interconnect is a crucial network technology for supercomputers. 
To support exascale systems, network resources are shared such that links and routers are not dedicated to any node pair.
While link utilization is increased, workload performance is often offset by network contention. 
Recently, intelligent routing built on reinforcement learning demonstrates higher network throughput with lower packet latency.
However, its effectiveness in reducing workload interference is unknown. 
In this work, we present extensive network simulations to study multi-workload contention 
under different routing mechanisms, intelligent routing and adaptive routing, on a large-scale Dragonfly system.
We develop an enhanced network simulation toolkit, along with a suite of workloads with distinctive communication patterns.  We also present two metrics to characterize application communication intensity.  Our analysis focuses on examining how different workloads interfere with each other under different routing mechanisms by inspecting both application-level and network-level metrics.  Several key insights are made from the analysis.

% High-radix Dragonfly interconnect is a crucial network technology for current and future supercomputers. 
% To support the unprecedented system scale, network resources in Dragonfly are shared such that links and routers are not dedicated to any node pair. While link utilization is increased, workload performance is often offset by network contention. Several approaches are presented for mitigating workload interference on Dragonfly.  Recently intelligent routing built on reinforcement learning is proposed for higher network throughput with lower packet latency. However, its effectiveness for reducing workload interference is unknown. In this work, we present extensive network simulations to study multi-workload contention 
% under different routing mechanisms --- intelligent routing and adaptive routing ---  on a large-scale Dragonfly system. We develop an enhanced network simulation toolkit, along with a suite of workloads with distinctive communication patterns. We also present two metrics to characterize application communication intensity. Our analysis focuses on examining how different workloads interfere with each other under different routing mechanisms by using both application-level and network-level metrics. Several key insights are made from the interference analysis.
\end{abstract}


\section{introduction}

High-performance computing (HPC) systems rely on efficient and scalable interconnect networks to support unprecedented system size at a reasonable cost.
Recent high-radix, low-diameter Dragonfly topology has demonstrated its high capability on the current Top500 list \cite{08df}\cite{top500}.
The next generation Slingshot interconnects also adopt Dragonfly topology to support future exascale HPC systems \cite{sc20slingshot}.

Dragonfly topology achieves extreme performance by arranging network resources of routers and links into fully connected groups.
As a result, Dragonfly is a \emph{hierarchical topology} that uses global links for inter-group connection and local links to connect routers in the same group.
The hierarchical design makes Dragonfly a low-diameter system such that any point in the network can be accessed by crossing the interconnect hierarchy in three hops. 
The fully connected inter-group structure also makes Dragonfly a \emph{high path diversity} network with abundant routing path possibilities.
A packet can either be minimally forwarded from its source group to the destination group or be relayed at an intermediate group in the system. 
Larger systems with more groups further increase the degree of path diversity.
While high path diversity provides more flexibility in choosing the packet forwarding path, it poses a significant challenge in routing design. 

%\subsection{Motivation}
Due to the sharing nature of Dragonfly, traffic flows from different applications have a high probability of competing for a portion or the entire network forwarding path, thus leading to \emph{workload interference}. 
Adaptive routing is widely deployed on production Dragonfly systems to deal with high path diversity and network congestion \cite{sc20slingshot,crayxcnetwork,sc12cray}. 
When forwarding a packet, adaptive routing \cite{08df,isca09indirectadp} chooses a path between either a minimal path or a non-minimal path via an intermediate group according to local information such as router port queue occupancy.
%However, local queue occupancy can only gauge near-end congestion but fails to estimate network condition at a few hops away \cite{dffarendcongest}.
%This inaccuracy often leads to sub-optimal network performance.
%The use of local queue occupancy in adaptive routing can further exacerbate network congestion and lead to severe workload interference. 
%However, local queue occupancy can only gauge near-end congestion but fails to estimate network condition at a few hops away \cite{dffarendcongest}.
%This inaccuracy often leads to sub-optimal network performance.
Although heuristic based adaptive routing can mitigate the workload interference in some degree, several empirical studies conducted on production Dragonfly systems have indicated that HPC applications can see a slowdown as much as 70\% \cite{sc17run2run, boyang-ROSS}.
Communication interference among different workloads can lead to many negative impacts, e.g., longer application execution time and wasted resource cycles, all of which lead to low system productivity. 

%\subsection{Limitation of state-of-art approaches}
In order to address workload interference, various approaches have been studied. 
Several efforts have demonstrated that careful job placement such as \emph{contiguous placement} can help mitigate workload interference \cite{ipdps20xin, sc16yang}.
Contiguous placement helps isolate an application from others by placing application processes in the same group.
%As a result, most messages are intra-group without the need to go through a global link.
While contiguous placement is shown to be effective for reducing workload interference through network simulation studies, it may cause local hot spots \cite{sc14jain}. 
In addition, contiguous placement is impractical in practice as it can cause severe system fragmentation.
For example, external fragmentation occurs when there is a sufficient number of compute nodes available for a job; however, they cannot be allocated to the job because these compute nodes are not in a contiguous partition.
Another research approach is \emph{application-aware routing} \cite{sc19adpthreshold}. 
The main idea is to build an analytical model to estimate network congestion and use it to adjust adaptive routing bias at the application's runtime. 
Continuous network monitoring is needed for performance modeling, which causes undesirable overhead to application performance.
Besides, the model may not be accurate which further impacts the effectiveness of this approach. 
Other more sophisticated methods such as \emph{congestion control}, with support of additional hardware and/or software, are also proposed to be used in combination with adaptive routing algorithms \cite{ics21qos,sc20slingshot}.
%However, they require additional hardware and/or software supports.

%\subsection{Key insights and contributions.}
%Q-adaptive routing is recently proposed for Dragonfly topology based on reinforcement learning techniques \cite{q-adp}.
%It shows superior performance than existing adaptive routing algorithms according to extreme traffic pattern analysis (e.g., well balanced network traffic and extremely imbalanced network traffic). 
%However, whether it is effective for reducing workload interference on Dragonfly systems is an open problem. 
In this study, we investigate workload interference by presenting \emph{a quantitative analysis of intelligent routing for mitigating workload interference on Dragonfly systems}. Q-adaptive routing is an intelligent routing method built on reinforcement learning for Dragonfly \cite{q-adp}. Preliminary results indicate that Q-adaptive routing provides superior performance than  adaptive routing under either well balanced or extremely imbalanced network traffics. However, its effectiveness for reducing workload interference is an open problem. 

\emph{Quantitative analysis of workload interference under routing on Dragonfly is challenging}. Dragonfly is a proprietary network technology, and is targeted for large-scale supercomputers such as those listed on the Top500 list. 
Although dozens of Dragonfly systems are available at national supercomputer centers, routing configuration is part of system configuration, which is impossible for general users to make changes at will \cite{sc14jain, sc11-abhinav,sc16yang}.  Following the common practice in the literature \cite{pmbs21-neil, cluster17-misbah}, 
we use high-fidelity flit-level network simulation in this study. 

 %(1) sst enhancement (2) app development (3) interference analysis (metrics?) and (4) key insights
Our study makes \emph{four contributions}. \textbf{First,} we enhance the well-known Structural Simulation Toolkit (SST) toolkit for the quantitative analysis (Section III). Several SST core and element modules are extended to support workload interference analysis at the flit level. The extended SST toolkit is available as open source software on GitHub \cite{sc_artifact}. 
\textbf{Second,}
in order to represent realistic HPC workloads, nine applications with distinct communication patterns are selected and developed for the study. These applications include both conventional HPC workloads and emerging machine learning workloads. 
They are also available on GitHub, along with the enhanced SST toolkit. Moreover, we present two metrics to formally characterize communication intensity of HPC applications (Section IV). We use the augmented SST toolkit to systematically analyze workload interference 
through pairwise workload and mixed workload analysis on a 1,056-node Dragonfly system.
\textbf{Third,}
even with high-fidelity network simulation, how to effectively study workflow interference with different routing mechanisms is quite challenging. In particular, an overwhelming amount of application-level and network-level data series can be collected. 
%workload interference is highly dynamic, and how to effectively analyze workload interference on Dragonfly is an open question. 
In this study, we propose two sets of quantitative metrics for interference analysis: (1) \emph{application-level metrics} such as application communication throughout, average application packet latency, and application communication time; and (2) \emph{network-level metrics} such as router port stall time, link traffic size and congestion index (Section \ref{sec:pairwise}-\ref{sec:mixedworkload}). These metrics are used in various simulation configurations to examine when and how interference occurs. 
\textbf{Fourth,}
several key  insights are obtained through the analysis, some being listed below:

\begin{itemize}

\item Application communication intensity can be well captured by two metrics, that it, message injection rate and peak communication ingress volume. An application with a lower message injection rate tends to be interfered by those with a higher rate, and an application with high peak ingress volume can cause severe workload interference to other applications.
%Although message injection rate is usually used to describe an application's network requirement, it is not the only metric that contributes to an application's communication intensity. Our study indicates that a new metric named \textit{peak ingress volume}, representing an application's maximum instant bandwidth requirement,  is especially useful for capturing the intensity of one-to-many communication.

\item Compared to adaptive routing, Q-adaptive routing can greatly reduce workload interference by saving up to 42.63\% communication time and mitigating up to 70.80\% communication performance variation caused by network contention. 
System-wide network analysis shows that Q-adaptive routing optimizes overall network utilization by balancing system traffic, eliminating hot spots and hence avoiding network congestion.
%We believe that intelligent routing such as Q-adaptive routing is orthogonal to the research of job placement and congestion management for tackling workload interference on Dragonfly systems.

\item For collective communication operations, tail latency control becomes important since the operation completes only after all have received the messages. Moreover, workload interference can be hidden by the application with long computation time and short communication time. 

\end{itemize}

%For the rest of the paper, Section \ref{sec:bg} presents the background with related work, Section \ref{sec:sim_design} discusses the detailed network simulation design followed by the studied applications in Section \ref{sec:apps}. 
%Section \ref{sec:pairwise} and \ref{sec:mixedworkload} provide pairwise and mixed workload interference analysis resepectively. Finally, Section \ref{sec:conclusion} concludes the paper.

\section{background and Related Work} \label{sec:bg}

\subsection{Dragonfly Topology}

\begin{figure}[ht]
  \centering
  \includegraphics[scale=\figscale]{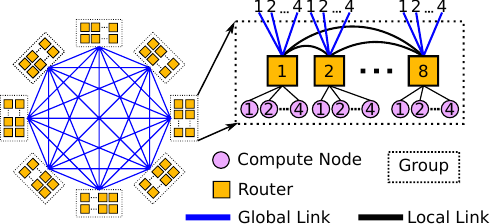}
  \caption{The hierarchical structure of 1,056-node Dragonfly system.}
  \label{fig:1ddf}
\end{figure}

Dragonfly is a \emph{hierarchical topology} by arranging network resources of routers and links into high-radix groups \cite{08df}. An example is shown in Figure \ref{fig:1ddf}.
The groups are fully connected by global links and routers in the same group are connected through local links.
Although groups are always all-to-all connected to assure system connectivity, the intra-group connection structure is not fixed and can have various forms.
For example, routers within the same group can also be all-to-all connected for higher system throughput or be arranged into a two-dimensional array or a tree to support a larger system size \cite{sc12cray,megafly,df+}.
In this study, we focus on the Dragonfly with a fully connected intra-group connection since this structure will be deployed on the next-generation HPC systems \cite{sc20slingshot}.

With a fully connected inter-group and intra-group structure, Dragonfly is a low-diameter topology.
Any packet can be delivered within three hops by crossing: (1) one local link in the source group to reach the router that has global link connection to the destination group, (2) one global link crossing groups, (3) and one local link in the destination group to the destination router.
Since two local links are required with one global link, local and global links are typically provisioned in a ratio of 2:1 on each router to make sure global link bandwidth is not wasted due to the insufficient local links \cite{08df}.

\subsection{Routing on Dragonfly}

Unlike fat-tree topologies, the hierarchical design and fully shared network resources make Dragonfly routing unique.
Static routing such as minimal routing that performs well on other topologies is not ideal for Dragonfly due to the limited number of global links between source and destination groups.
%Statically forwarding packets often leads to unbalanced network usage.
Hence, Dragonfly relies on adaptive routing to let routers sense local congestion and bypass hot spots by forwarding packets non-minimally through an intermediate group if necessary.
%Since groups are all-to-all connected, any group within the system can be chosen as the intermediate, resulting in a large number of candidate forwarding paths. 
%High path diversity makes Dragonfly routing more flexible, however, it is also very challenging to choose the best path.
\textbf{Adaptive routing} is a class of heuristic methods that use router's port queue occupancy to estimate network congestion.
When forwarding a packet, a router first randomly selects two minimal paths and two non-minimal paths and checks their corresponding port queue occupancy \cite{crayxcnetwork}.
When the best minimal path queue occupancy is less than twice of the best non-minimal path queue occupancy, the packet is minimally forwarded, otherwise, the non-minimal path is used.

Depending on where the dynamic routing decision is made, adaptive routing algorithms have three widely used solutions:
Universal Globally-Adaptive Load-balanced routing (UGAL) lets the source router make a one-time routing decision with two variants.
In case a packet enters an intermediate group, \textbf{UGALg} (UGAL\_group) forwards it minimally towards the destination group, whereas \textbf{UGALn} (UGAL\_node) tries to avoid local link congestion by having the packet firstly visit a random router in that intermediate group \cite{dffarendcongest}.
\textbf{PAR} (Progressive Adaptive Routing) is similar to UGALn, except that it allows routers in the source group to revise the previously made minimal routing decision in case local congestion is encountered at downstream routers \cite{isca09indirectadp}.
Although adaptive routing is good at avoiding near-end congestion, they are oblivious to the network condition at a few hops away hence may cause network congestion.

\begin{figure}
    \centering
    \includegraphics[scale=\figscale]{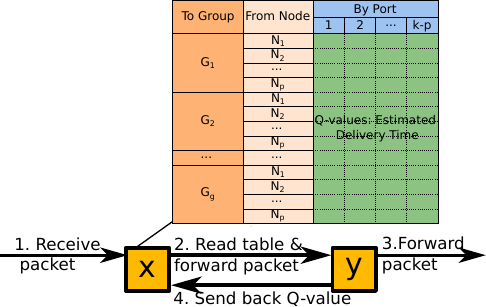}
    \caption{Q-adaptive routing and its two-level Q-table per router. Router X follows four steps to forward a packet and update its table. Through the process of sending packets and receiving feedback signals from the neighbors, every router learns the system-wide network condition and records the knowledge in its two-level Q-table for packet forwarding.
    %When router X needs to forward a packet, its Q-table, which stores the learned network information, is firstly consulted to select the most efficient forwarding path.
%After forwarding the packet, router X update its Q-table by receiving a feedback signal from its downstream router Y, who is closer to the packet destination and thus has more accurate network condition towards that point. 
}
    \label{fig:q-adp-flow}
\end{figure}

\textbf{Q-adaptive Routing} is a recent Dragonfly routing solution leveraging reinforcement learning (RL) techniques \cite{q-adp}. In Q-adaptive routing, each router learns the overall network condition, records the knowledge in a light-weight two-level Q-table, and uses the table for packet forwarding. Through the process of sending packets and receiving feedback signals from the neighbors, every router learns the system-wide network condition and stores the information in the Q-table. Similar to adaptive routing algorithms, Q-adaptive routing forwards packets either minimally or non-minimally. 
In other words, unlike adaptive routing which makes routing decisions based on local information, Q-adaptive routing uses \emph{the learned overall system condition and global path congestion} to direct packet delivery (Figure \ref{fig:q-adp-flow}). In addition, Q-adaptive routing is fully distributed such that routers make their independent decisions without requiring any shared information. 

%Second, unlike reinforcement learning built upon deep neural networks, table-based Q-adaptive routing is easy to implement and much more responsive thanks to the short table look-up time.
%\ul{**yao: I feel like cold start is not a feature of Q-adp, Deep Q learning can replace Q-table with NN and can also start from a cold system.
%We start from a cold system to make the comparison more fair and results more convincing. 
%Q-adaptive routing can be used with cold start (i.e., to give recommendation with an initial table filled with random numbers which will be updated as more data is available).} 
%The table based, fully distributed nature make Q-adaptive routing a practical solution for Dragonfly systems that can be easily implemented without any additional hardware.
%In \cite{q-adp}, it is shown that Q-adaptive routing outperforms adaptive routing under well-balanced and extremely imbalanced traffic patterns.

In this study, we investigate workload interference under the aforementioned routing mechanisms on Dragonfly systems. 
\subsection{Related Works}

Chunduri et al. studied Dragonfly network interference on production systems and observed as much as 7x performance variability for MPI collectives and 70\% variability for MILC application due to the network contention \cite{sc17run2run}.
Simulation studies show that network interference also appears on Dragonfly variants under different intra-group connection structures \cite{ipdps20xin, pads19kang}.
Yang et al. observed the bully effect on the Dragonfly system that communication intensive application affects less intensive applications and purposed using contiguous placement to reduce network interference \cite{sc16yang}.
Jain et al. proposed to use random placement to improve system throughput by reducing network hot-spots on Dragonfly systems \cite{sc14jain}.
In addition to job placement, QoS is another approach to mitigate network interference by separating traffic flows of different applications or communication types into isolated channels \cite{ics21qos,misbah19qos,cluster20qos}.
At the routing level, Smith et al proposed adaptive flow-aware routing to mitigate network interference \cite{sc18adprting}.
De Sensi et al proposed to use dynamic adaptive routing bias to constrain the contention effect \cite{sc19adpthreshold}.
A more complicated congestion control mechanism is also proposed such that when congestion happens, the message generation rate is throttled to drain the network system \cite{sc20slingshot, pmbs21-neil-cc}.

%The existing studies either focus on job placement or enhancing adaptive routing algorithms. %Unlike the aforementioned studies, our work explores reinforcement learning routing for workload interference migration on Dragonfly systems. 
%By using a set of representative applications with distinctive communication patterns observed in both conventional scientific applications and emerging machine learning applications, we examine Q-adaptive routing versus the existing adaptive routing methods under a variety of pairwise application analyses as well as mixed workload analysis. 
This work presents a quantitative analysis of 
different routing methods with respect to workload interference mitigation, hence complementing the above related interference studies.

\section{Network Simulation Design} \label{sec:sim_design}

Workload interference can be analyzed through application tracing and profiling. However, the data collected in the trace is limited to the given application, and the information of other co-running applications is not included. 
Although analyzing application trace or profile can unveil the interference effect of the studied application, it fails to answer how and when workload interference occurs. 
%Additionally, system-wide hardware counters are not always available to the researchers, and the unpublished detail of the routing implementation on real systems makes it difficult to reason the interference root cause. 

High-fidelity network simulation is a viable approach to study workload interference \cite{mubarak2016-codes, sc16yang}. In this study, we leverage the well-known Structural Simulation Toolkit (SST) for interference analysis \cite{sst}. The SST toolkit contains a number of core and element libraries. 
In this work, we enhance three SST libraries, namely Ember, Firefly, and Merlin, for the quantitative network study.
\textit{SST/Ember} contains a collection of MPI motifs that represent real-world application communication behaviors.
The communication generated by SST/Ember is passed to \textit{SST/Firefly}, a state machine based MPI implementation layer that simulates blocking, non-blocking communication, and MPI collective operations with eager and rendezvous protocols.
Finally, the Firefly MPI messages are packetized and transmitted to \textit{SST/Merlin}, which provides high-fidelity, flit-level network simulation including Dragonfly topology. 

\begin{figure}
    \centering
    \includegraphics[scale=\figscale]{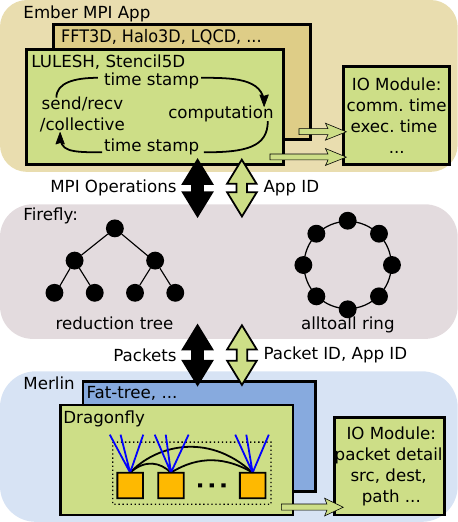}
    \caption{Enhancing SST for workload interference analysis. Our enhancements are shaded in green.}
    \label{fig:sst}
\end{figure}

Figure \ref{fig:sst} depicts the interplay of these SST libraries, along with our enhancements. 
Specifically, we implement an IO module that can record any desired performance counter at any frequency.
For the purpose of simulation efficiency, the IO module can be flexibly configured to coalesce multiple write operations into one action to balance the trade-off between IO efficiency and system memory usage.
The designed IO module enables us to investigate every detail of a simulated system at both application and network level.
At the application level, all the studied MPI applications are updated to accurately timestamp each iteration's starting, ending time, and time spent on different messaging operations. The applications are connected with the IO module to record all the metrics for post-simulation analysis.
We also enhance SST/Merlin Dragonfly topology by connecting it with the IO module to record every packet’s detailed information including source,
destination, sending, receiving time, and forwarding path. 
Moreover, we enhance the bridge
between SST/Ember and SST/Merlin by passing application’s
job ID to provide fine-grained per-application network link
usage statistics. 

%The main challenges in studying Dragonfly network interference are due to its highly dynamic and quickly evolving nature.
%With adaptive routing algorithms, the packet forwarding path is determined according to the network condition at its specific moment, which changes instantly.
%The highly dynamic nature makes it difficult to capture and analyze intermittent network congestion as it may disappear quickly due to the routing adjustment and then reappear at different locations.

%Previous simulation studies analyze network interference with the application communication traces captured from real executions. 
%Although traces represent applications at a more realistic level, trace replaying is often very expensive with high computation and memory requirement. 
%Moreover, due to the limited number of captured applications, trace replaying fails to represent the shared HPC systems with various types of applications and loses the simulation flexibility due to the fixed job size given by the trace.

In this study, we focus on \emph{a 33-group, 1,056-node Dragonfly system} as shown in Figure \ref{fig:1ddf}. 
Each group contains eight fully connected routers with a total of 32 global links.
Each router hosts four compute nodes using 128B flits and 512B packets. 
To ensure network bandwidth is not limited by the flying credits of credit-based flow control, each router port can buffer up to 30 packets. Local and global links are configured with 200Gb/s bandwidth according to the Slingshot system \cite{sc20slingshot}. 
Considering the difference in link length, flit transmission latency is 30ns and 300ns for local and global links to keep the 1:10 ratio  as used in previous works \cite{08df,isca09indirectadp,dffarendcongest}. 
The evaluated adaptive routing algorithms are configured with zero bias towards the minimal path and Q-adaptive routing with the same hyperparameters as in \cite{q-adp}.
%For the studied applications, we focus on its network performance by placing one process per node, and its placement is randomly selected following the behavior of current scheduling policies on production systems.

\section{Applications and Communication Patterns} \label{sec:apps}

Although existing studies show that communication-intensive applications can bully other applications on Dragonfly \cite{sc16yang}, \emph{there is no formal definition of communication intensity}. We propose two metrics to characterize application communication patterns. Both metrics can be measured or be derived from the application source code.

\begin{itemize}
\item  \textit{Message injection rate} is defined as an average value derived from an application's total message size and its execution time. It is measured as the average bandwidth requirement of an application assuming its packets are steadily injected into the network. 
\item \textit{Peak ingress volume} is defined as an application's peak bandwidth requirement, meaning the amount of message that the network is expected to process in a short time.
It is measured as the consecutive message size injected into the network by the application. 
%This metric is determined by the message size as well as the number of communication partners in the cases of one-to-many communication patterns.  
\end{itemize}

\begin{table}[!t]
\caption{List of applications with their communication patterns.}
\begin{center}
\renewcommand{\arraystretch}{1.3}
\begin{tabular}{
>{\raggedright}p{0.1\linewidth}
>{\raggedright}p{0.1\linewidth}
>{\raggedleft}p{0.14\linewidth}
>{\raggedleft}p{0.13\linewidth}
>{\raggedleft}p{0.15\linewidth}
>{\raggedleft\arraybackslash}p{0.12\linewidth}}
%\begin{tabularx}{\linewidth}{XXXXXX}
\toprule
Pattern   & App       & Total Msg (MB) & Execution time (ms) & Injection Rate (GB/s) & Peak Ingress Volume \\ %$^{\mathrm{*}}$
\midrule
Random    & UR        & 11829.48      & 13.31        & 888.48           & 3.07KB             \\
\hline \hline
Sweep     & LU        & 13713.22      & 13.71        & 999.88           & 30.0KB               \\
Alltoall  & FFT3D     & 15781.09      & 12.53        & 1259.35          & 51.68KB            \\
Stencil   & Halo3D    & 47769.10      & 10.85        & 4403.81          & 1.15MB             \\
          & LQCD      & 11924.31      & 13.79        & 864.70           & 4.60MB              \\
          & Stencil5D & 9833.95       & 13.70        & 717.87           & 14.0MB             \\
\hline \hline
Allreduce & CosmoFlow & 2373.84       & 13.65        & 173.86           & 2.25MB             \\
          & DL        & 9714.44       & 11.86        & 819.12           & 2.30MB            	\\
\hline \hline
Stencil   & \multirow{2}{*}{LULESH}   & \multirow{2}{*}{17900.12} & \multirow{2}{*}{12.34} & \multirow{2}{*}{1450.78} & 1.95MB \\
+Sweep    &                         &                           &                        &                          & 14.91KB \\
%stencil   & \multirow{2}{*}{LULESH} & \multirow{2}{*}{17900.12} & \multirow{2}{*}{12.34} & \multirow{2}{*}{1450.78} & 1.95MB             \\
%+sweep    &                         &                           &                        &                          & 14.91KB           \\
\bottomrule
\end{tabular}
\label{tab:app_pattern}
\end{center}
\end{table}

In order to well capture representative workloads on production systems, we carefully select nine applications based on their communication patterns. They cover a broad range of workloads, which are grouped into five representative communication patterns commonly observed in the traditional scientific and the emerging machine learning applications. Some applications are from the SST package, and some are developed in this work. Table \ref{tab:app_pattern} summarizes these applications, along with their communication patterns. 

\begin{itemize}

	\item \textbf{Random}:
	Random is a typical one-to-one communication pattern in the scientific field to study network performance.
	The Uniform Random (\textit{\textbf{UR}}) is mainly used as a background application to mimic a system under a balanced network load with each process sending messages to random targets.
			
	\item \textbf{Sweep}: 	
	The Lower-Upper (\textbf{\textit{LU}}) Gauss-Seidel solver application from the NAS Parallel Benchmark suite is featured by a sweep communication pattern \cite{naslu}. 
	LU arranges its processes as a 2D square and initiates the network communication from one square corner.
	All processes send data to their downstream neighbors after receiving messages from their upstream peers along both square dimensions.
	As a result, the communication behaves like a wavefront sweeping from one corner of the square to the other diagonal corner stage by stage, which causes long communication latency.
	Because each process has two downstream partners, LU's peak ingress volume counts two messages.
	
	\item \textbf{Alltoall}: 
	This MPI collective is important for applications that perform parallel Fast Fourier Transforms (FFT) such as pF3D \cite{pf3d}, NAMD \cite{namd}, and VASP \cite{vasp}.
	SST implements MPI\_Alltoall through a multi-step ring exchange such that process $N$ receives data from process $N-i$ and sends data to process $N+i$ in round $i$.
	Since each process only sends one message in each round, Alltoall operation's peak ingress volume only counts one message.
	When the communicator is large, many rounds of send and receive are required hence leading to long communication latency.
	
	The studied \textbf{\textit{FFT3D}} application maps the problem to a two-dimensional processes array, where processes on the same row and column perform Alltoall operations.
	
	\item \textbf{Stencil}: 
	Stencil computation is an important class of algorithms in scientific computing and a representative one-to-many communication pattern \cite{stencil1}.
	Under a stencil pattern, processes are arranged into a multi-dimensional grid, and each process talks with its nearest neighbors in both directions along all the dimensions. 
	As a result, stencil's peak ingress volume must consider both the message size and process grid dimension.
	Because at each communication step, processes send data to multiple destinations, stencils are high-throughput and aggressive on occupying network resources.
	
	\textit{\textbf{Halo3D}} is a 3D-stencil application that processes communicate with its six neighbors along three dimensions.
	Lattice Quantum ChromoDynamics (\textbf{\textit{LQCD}}) is a QCD simulation application to study strong force theory featured with a 4D-stencil pattern of eight neighbors \cite{lqcd}.
	\textit{\textbf{Stencil5D}} is a synthetic application used to study the impact of peak ingress volume that processes have up to ten neighbors. 

	\item \textbf{Allreduce}: 
	Distributed machine learning applications periodically aggregate and distribute the learned information among all processes to update model parameters \cite{horovod}.
	As a result, collective communication such as Allreduce becomes heavily used and is presented as an important message movement pattern on the HPC systems.
	In SST, MPI\_Allreduce is implemented by arranging processes into a binary tree where the desired data is firstly aggregated from leaf to root and then distributed in the opposite direction.
	Allreduce peak ingress volume counts two messages considering each tree node has two children. 
	
	\textbf{\textit{CosmoFlow}} is a fully synchronous data parallel distributed deep learning (DL) cosmology application \cite{sc18cosmoflow}.
	CosmoFlow shows strong scaling performance with the main communication pattern of 28.15 MB Allreduce messages every 129 ms.
	To make the application have similar execution time as other applications, we proportionally decrease both message size and communication interval by 25x and keep its intrinsic communication intensity.
	To represent large-scale distributed DL applications with a massive training dataset, a heavier Allreduce application named \textbf{\textit{DL}} is studied.
	DL has similar message size but shorter communication interval, such that its message injection rate is around 4.7x higher than that of CosmoFlow. 
	
	\item \textbf{Hybrid}:
	The Livermore Unstructured Lagrangian Explicit Shock Hydrodynamics (\textbf{\textit{LULESH}}) is a widely studied proxy application designed for exascale hardware/software co-design effort \cite{ipdps13lulesh}.
	LULESH represents a typical hydrocode such as ALE3D using MPI non-blocking (MPI\_Isend, MPI\_Irecv) and collective operations \cite{LULESH2:changes}.
	LULESH communication pattern is dominated by a 26-point 3D-stencil followed by a sweep3D data exchange \cite{pads17lulesh,hpdc15lulesh}.
	LULESH shows good weak scaling performance on production systems \cite{ipdps14lulesh}. 
	We implement LULESH based on the communication analysis listed in the literature \cite{pads17lulesh, hpdc15lulesh}.

\end{itemize} 

%As we will show later, these metrics can well capture application communication intensity.

%For each application, in addition to the total message size and execution time, communication intensity is characterized by two additional metrics: \textit{message injection rate} and \textit{peak ingress volume}.

%A higher peak ingress volume depicts a more aggressive network bandwidth requirement in a short time to absorb all the messages into the network.

%Among these applications, CosmoFlow has the least message injection rate, whereas stencil application such as Halo3D has the largest injection rate and Stencil5D has the highest peak ingress volume.

%We quantitatively study the network workload interference through pairwise workload analysis and mixed workload analysis.

\begin{figure*}[ht]
  \centering
  \includegraphics[scale=\figscale]{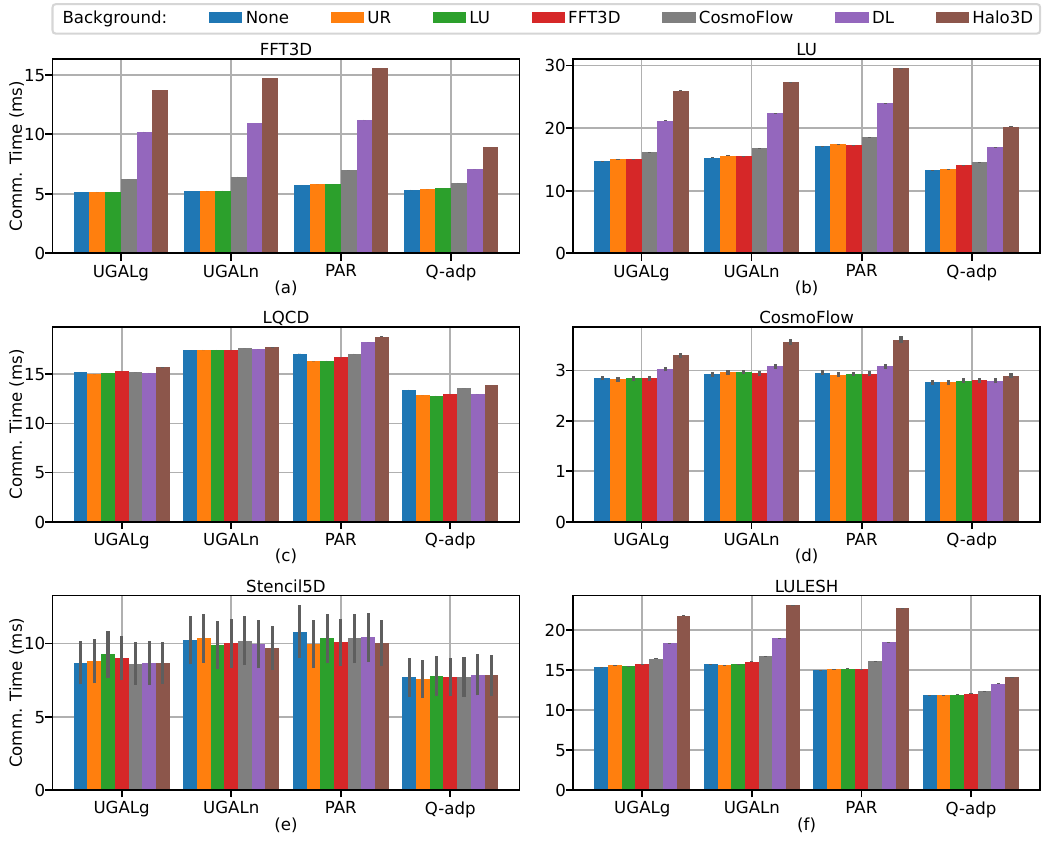}
  \caption{Average communication time (bar) and standard deviation (vertical line) of a target application over all processes.
  The results for six target applications are presented in (a)-(f). For a target application, each colored bar indicates its communication time under a background application (or none). 
%   and the variation of communication time shows the effect of network interference.
  }
  \label{fig:random_app}
\end{figure*}

\section{Pairwise Workload Analysis} \label{sec:pairwise}

In this section, we study pairwise workload interference by co-running \emph{a target application} with \emph{a background application}. The goal is to analyze how the communication performance of the target application is impacted by the background application with different message injection rates and peak ingress volumes.
Specifically, FFT3D, LU, LQCD, CosmoFlow, Stencil5D and LULESH are selected as the target applications covering different communication patterns and intensities.
Each target application is co-run with a background application from different communication pattern categories.
For pairwise analysis, the 1056-node system is equally divided to host a target application and a background application.
To maintain a perfect process cube, LULESH takes 512 (=$8^3$) nodes leaving 16 nodes being idle.
% Other target and background applications occupy all the half of system nodes.
Random job placement is used in our experiments.
%Figure \ref{fig:random_app} summarizes target applications' communication time under different background applications, where the bars show the average communication time and ``\verb|X|'' marks the maximum.

For each target application, its process-to-node mapping is kept unchanged across different runs (i.e., standalone or co-running with a background application).
Therefore, a variation in communication time of the same application under different runs unveils the effect of workload interference. 
\emph{To make a fair comparison, Q-adaptive starts an application under the same condition as adaptive routing algorithms without any pre-trained information.}
Thus, the time spent on training Q-adaptive routing is included in its communication time.

Note that we use both \emph{application-level metrics and network-level metrics} for interference analysis. 
We mainly focus on the communication time and its variation across application processes for the application-level analysis.
We leverage packet latency distribution, tail latency dispersion, and application communication throughput for the network-level analysis.
%We quantify the severity of workload inference by evaluating the average and maximum communication time over all processes of the target applications. 
%The communication time is measured as the wall clock time of a process spent on sending and receiving messages without taking its computation into account.

%\begin{tcolorbox}
%Q-adaptive outperforms adaptive routing algorithms when an application is the only job on the system.
%\end{tcolorbox}

In order to clearly identify workload interference, we first conduct standalone experiments to collect target application performance in an interference-free environment.
%These results will be presented throughout the rest of the paper to show workload interference.
Figure \ref{fig:random_app} presents the application communication time under different pairwise analyses. 
The blue bars show the communication time of applications when it is the only job on the system.
For the standalone cases, FFT3D and CosmoFlow perform similarly among different routing algorithms, 
whereas LU, LQCD, Stencil5D, and LULESH perform the best under Q-adaptive routing with an average of 23.46\% smaller communication time compared with PAR.

In short, \emph{compared with adaptive routing, Q-adaptive achieves equal or better performance when each target application runs alone with random placement.}

%The blue bars in Figure \ref{fig:random_app} show applications' communication time when it is the only job on the system.
%In Figure \ref{fig:random_app}(a), FFT3D performs similarly under adaptive and Q-adaptive routing algorithms.
%The best performing one is UGALg with a 5.13ms average communication time, whereas that of PAR is 5.70ms. 
%Q-adaptive routing achieves 5.35ms average communication time, which is 4.29\% longer than UGALg but 6.23\% better than PAR.

%Q-adaptive has the best performance for all the other target applications.
%As shown in Figure \ref{fig:random_app}(b), LU has the smallest average communication time of 13.24ms using Q-adaptive routing, which is 10.51\%, 13.36\%, and 22.64\% smaller than that of UGALg, UGALn, and PAR respectively.
%Similarly, in Figure \ref{fig:random_app}(c), the average communication time of LQCD with Q-adaptive routing is 13.43ms, which is 11.74\%, 22.95\%, and 21.49\% smaller than UGALg, UGALn, and PAR respectively.
%In Figure \ref{fig:random_app}(d), CosmoFlow spends an average of 2.76ms for communication, which is still the smallest compared with 2.92ms, 2.96ms, and	2.76ms when using UGALg, UGALn, and PAR.

\begin{figure*}[ht]
  \centering
  \includegraphics[scale=\figscale]{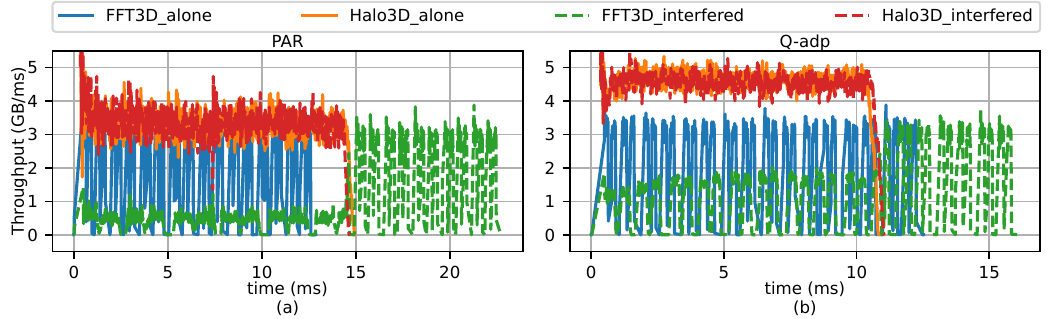}
  \caption{Network throughput of FFT3D and Halo3D along simulated time. Q-adaptive protects FFT3D's performance from Halo3D's interference with 2.58x higher throughput compared with that of PAR shown in green lines.}
  \label{fig:random_fft_halo_bw}
\end{figure*}

\subsection{Impact of Message Injection Rate}

\begin{tcolorbox}
Applications with a higher injection rate tend to delay those with a lower rate.
%Communication intensive applications tends to affect less intensive ones. 
Q-adaptive routing reduces workload interference with shorter communication time and less variation.
\end{tcolorbox}

Figure \ref{fig:random_app}(a) shows that UR or LU causes negligible interference effect on FFT3D. 
The average communication time of FFT3D in both cases increases by less than 3\% with all the studied adaptive and Q-adaptive routing algorithms.
As summarized in Table \ref{tab:app_pattern}, FFT3D is hardly affected by UR and LU, both having lower injection rates.
CosmoFlow causes a mild interference impact on FFT3D with an average of 22.51\% communication delay under adaptive routing and a smaller 10.81\% delay under Q-adaptive routing. 
Although the injection rate of CosmoFlow is smaller than that of FFT3D, CosmoFlow has a considerably larger peak ingress volume.
The effect of peak ingress volume is discussed in \S\ref{sec:peak ingress vol}.
Compared with CosmoFlow as background application, DL --- a higher injection rate application --- causes more severe interference impact on target applications.
Interference from DL makes FFT3D spend 2.00x, 2.11x, 1.97x, and 1.32x more time on communication with UGALg, UGALn, PAR, and Q-adaptive routing, respectively. 
Among adaptive routing algorithms, PAR's 11.24ms average communication time is slightly longer than UGALg's 10.23ms and UGALn's 10.98ms. 
However, PAR has a smaller communication time variation of 97.14\% compared with UGALg's 99.53\% and UGALn's 110.77\% variation.
Although network interference is inevitable, Q-adaptive has the shortest communication time with the most stable performance. 
Q-adaptive's 7.08ms communication time is 37.05\% smaller than that of PAR.
Compared with the standalone case, Q-adaptive's communication time variation is 32.34\%, which is 70.80\% smaller than UGALn's 110.77\% variation.
The highest injection rate application of Halo3D causes the most interference. 
Halo3D delays FFT3D's communication time by 2.67x, 2.83x, and 2.73x using UGALg, UGALn, and PAR respectively.
However, Q-adaptive reduces the interference effect by completing the communication within 8.93ms, which is 42.63\% smaller than PAR's 15.57ms.

To investigate the effect of injection rate, we examine the FFT3D-Halo3D co-running case.
Figure \ref{fig:random_fft_halo_bw} shows the application communication throughput along the simulated time under PAR and Q-adaptive routing. 
Both standalone and interfered cases are plotted in different colors.
The standalone results in Figure \ref{fig:random_fft_halo_bw}(a) and (b) show that Halo3D (orange line) is a communication intensive application with continuous high throughput. 
% Q-adaptive routes Halo3D packets more efficiently with an average of 4.55GB/ms throughput, which is 37.01\% higher than PAR's 3.32GB/ms. 
% Higher throughput delivered by Q-adaptive makes Halo3D run 27.45\% faster than PAR with 10.85ms execution time.
In contrast, FFT3D (blue line) features burden network throughput on which the valleys correspond to the forward and backward FFT computation and the peaks for the Alltoall operation between computations.
When FFT3D and Halo3D are co-run together, Halo3D (red line), as the application with a higher injection rate, is hardly affected by FFT3D, with less than 2\% variation with respect to both average throughput and application execution time under PAR and Q-adaptive routing.
However, Halo3D maintains its high network throughput at the cost of degraded FFT3D performance (green line).
In Figure \ref{fig:random_fft_halo_bw}(a), due to Halo3D's aggressive network requirement, FFT3D's average throughput (green line) using PAR is only 0.44GB/ms, which is 82.62\% smaller than the 2.51GB/ms throughput of standalone case.
Once Halo3D finishes at 14.66ms, FFT3D regains network bandwidth with an average of 2.46GB/ms throughput.
In contrast, Q-adaptive is capable of protecting FFT3D's performance while keeping Halo3D's high throughput.
In Figure \ref{fig:random_fft_halo_bw}(b), with Q-adaptive routing, the average throughput of FFT3D under Halo3D's interference is 1.13GB/ms, which is 2.58x higher than PAR's 0.44GB/ms.
Once Halo3D finishes at 11.05ms, FFT3D regains network resources with an average of 2.50GB/ms throughput.

A similar effect is also observed for the LU shown in Figure \ref{fig:random_app}(b).
UR and FFT3D cause no more than 6.06\% variation on average communication time for all the studied routing algorithms. 
CosmoFlow delays LU's communication by 9.34\% on average under adaptive routing and by 9.99\% under Q-adaptive routing. 
In the Allreduce pattern category, the higher injection rate DL application causes more impact on LU. 
When co-running with DL, LU spends 43.49\%, 46.63\%, and 40.16\% more time on communication under UGALg, UGALn and PAR, respectively.
PAR has a more stable performance with a slightly smaller communication time variation compared with other adaptive routing algorithms.
Nevertheless, Q-adaptive's 16.96ms communication time is the smallest among all the studied routing algorithms, with only a 28.15\% variation. 
The highest injection rate application, Halo3D, leads to significant interference on LU with the largest communication time and the greatest variation.
The three adaptive routing algorithms spend an average of 27.63ms on communication with an average of 75.77\% variation compared with the standalone cases.
Although Q-adaptive's average communication time is increased to 20.24ms from 13.24ms, the interference effect is still the smallest compared with adaptive routing algorithms.

Figure \ref{fig:random_app}(f) confirms the effect of message injection rate on workload interference. 
Significant interference effect is only observed with Halo3D background application, whose injection rate is the highest among all the studied applications.
Halo3D causes LULESH to spend an average of 46.81\% more time on communication under adaptive routing and only 19.62\% more time under Q-adaptive routing.
DL affects LULESH's communication in a much smaller magnitude with an average of 21.08\% delay under adaptive routing and only 12.30\% under Q-adaptive routing.
Other background applications cause no more than 7.35\% and 4.35\% communication time increase under adaptive and Q-adaptive routing respectively. 

\subsection{Impact of Tail Latency}

\begin{tcolorbox}
MPI collective performance degradation caused by network interference is manifested by long tail latency. 
\end{tcolorbox}

\begin{figure}[ht]
  \centering
  \includegraphics[scale=\figscale]{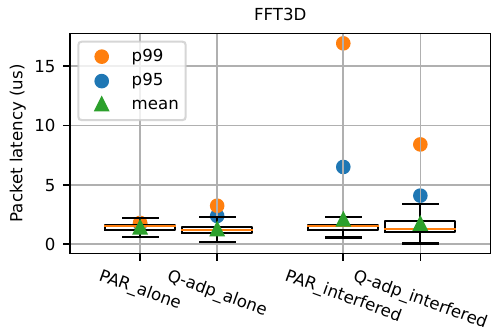}
  \caption{FFT3D packet latency distribution under both standalone and interfered by Halo3D cases with the 95th and 99th percentile latencies.}
  \label{fig:random_fft_halo_lat}
\end{figure}

In Figure \ref{fig:random_app}(a), Q-adaptive reduces up to 42.63\% communication time for FFT3D under the network interference from Halo3D. 
Q-adaptive has better performance thanks to its better packet tail latency control capability.
Figure \ref{fig:random_fft_halo_lat} depicts FFT3D's packet latency distribution with 95th and 99th percentile latency for both standalone and interfered by Halo3D cases.
The box ranges from the first quartile (Q1) to the third quartile (Q3) of the data, with a yellow line at the median.
% and a green triangle marking the mean.
When FFT3D is the only job on the system, Q-adaptive has a better average packet latency with a mean of \SI{1.27}{\micro\second} and a median of \SI{1.16}{\micro\second}, which are 10.32\% and 25.15\% smaller than those of PAR.
Because Q-adaptive routing starts without any pre-trained information, the learning and exploration process of reinforcement learning algorithm makes a few packets being routed with longer latency, shown as the larger tail latency under the standalone case.
% The 95th and 99th percentile latency under standalone case of Q-adaptive is \SI{2.37}{\micro\second} and \SI{3.23}{\micro\second}, which are larger than PAR's \SI{1.70}{\micro\second} and \SI{1.78}{\micro\second}.
% Q-adaptive has larger tail latency because when FFT3D starts to run, the learning and exploration process of reinforcement learning algorithm makes a few packets being routed with longer latency.
However, by comparing the standalone communication time in Figure \ref{fig:random_app}(a), this small difference in tail latency is negligible for overall communication performance.
% this less than \SI{1.5}{\micro\second} tail latency difference causes a negligible effect on the overall communication performance.

% \begin{figure}[ht]
%   \centering
%   \includegraphics[scale=1]{./exppdf/random_fft3d_halo3d_lat_logscale}
%   \caption{\textcolor{blue}{Same figure in log scale}}
%   \label{fig:random_fft_halo_lat}
% \end{figure}

% \begin{figure*}[ht]
%   \centering
%   \includegraphics[scale=0.9]{./exppdf/random_lqcd_halo3d_bw}
%   \caption{LQCD and Halo3D network throughput along simulated time. The more aggressive LQCD with a 4D-stencil pattern reduces Halo3D's throughput.}
%   \label{fig:random_lqcd_halo_bw}
% \end{figure*}

\begin{figure*}[ht]
  \centering
  \includegraphics[scale=\figscale]{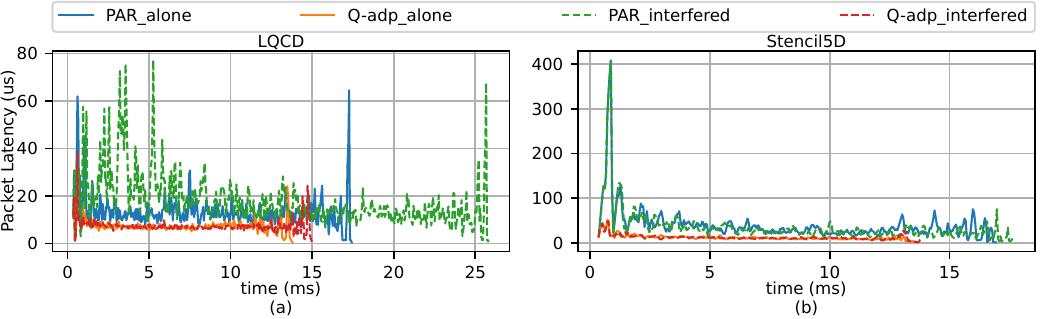}
  \caption{Packet latency of LQCD and Stencil5D. Stencil5D with larger peak ingress volume delays LQCD's packets significantly under PAR.}
  \label{fig:random_lqcd_stencil5d_lat}
\end{figure*}

When FFT3D is interfered by Halo3D, Q-adaptive routing provides much better tail latency control.
Although PAR and Q-adaptive have similar median packet latency (\SI{1.55}{\micro\second}, \SI{1.28}{\micro\second}), the 95th and 99th percentile latency of PAR are significantly delayed to \SI{6.50}{\micro\second} and \SI{16.92}{\micro\second}, which are 1.59x and 2.01x longer than Q-adaptive's \SI{4.08}{\micro\second} and \SI{8.40}{\micro\second}.  
For the Alltoall operation, it is only considered completed by a process when all messages are received.
Therefore, the tail latency has a more important influence on the application's communication time than the mean and median. 
As a result, we observe that compared with PAR, although Q-adaptive has similar median packet latency, its 50.35\% smaller 99th percentile latency helps FFT3D save 42.63\% time spent on communication as shown in Figure \ref{fig:random_app}(a) brown bars.

\subsection{Impact of Peak Ingress Volume} \label{sec:peak ingress vol}

\begin{tcolorbox}
Applications with larger peak ingress volume can better tolerate other application's interference by aggressively occupying network resources.
\end{tcolorbox}

According to Table \ref{tab:app_pattern}, the peak ingress volume of LQCD is larger than all the background applications. 
As shown in Figure \ref{fig:random_app}(c), LQCD is nearly immune to the network interference.
With PAR, even the highest injection rate Halo3D application only makes LQCD spend up to 10.10\% more time on communication.
Similarly, the largest peak ingress volume application, Stencil5D, experiences no more than 8\% communication time variation under all the background applications as shown in Figure \ref{fig:random_app}(e).
Stencil5D has higher communication time variance among its processes due to the imperfect multidimensional process cube.
% In effect, limited by the number of processes, 
It is difficult to create a balanced 5D-grid with similar size along five dimensions.
As a result, many processes are on the grid edges or surfaces with fewer neighbors, hence completing communication faster than those in the center of the grid. 

LQCD has a 4D-stencil communication pattern with maximum of 8 neighbors, and Stencil5D has up to 10 neighbors. 
Both are considered communication intense applications with large peak ingress volume. 
Since router buffers and links process packets on a first-come, first-served basis, applications with a higher peak ingress volume have a greater chance to insert their packets in front of other applications.
As a result, packets of applications with a smaller peak ingress volume, such as Halo3D, have to be queued and delayed.

To verify this effect, LQCD is co-run with Stencil5D, whose peak ingress volume is the largest among all the studied applications.
Figure \ref{fig:random_lqcd_stencil5d_lat} illustrates the effect of packet delay by plotting packet latency along simulated time.
In Figure \ref{fig:random_lqcd_stencil5d_lat}(a), when LQCD is the only job in the system, its mean and 99th percentile latency with PAR are \SI{13.59}{\micro\second} and \SI{42.94}{\micro\second} in the first 15ms.
However, introduction of Stencil5D into the system delays those values to \SI{21.37}{\micro\second} and \SI{77.44}{\micro\second}, a 57.32\% and 80.36\% increase respectively.
\begin{figure}[ht]
  \centering
  \includegraphics[scale=\figscale]{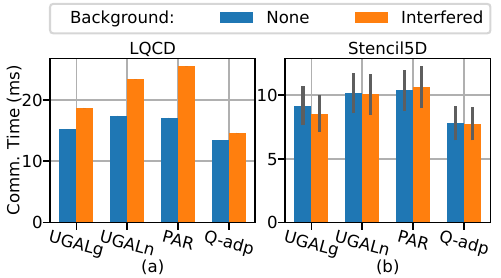}
  \caption{
  Communication time of LQCD and Stencil5D.
  ``None" means the application is the only job on the system and ``interfered" denotes the pairwise running of LQCD and Stencil5D. 
  LQCD is heavily interfered by Stencil5D, whose peak ingress volume is the largest.
  }
  \label{fig:commTime_lqcd_stencil5d}
\end{figure}
As expected, the significantly delayed LQCD packets force the application to spend more time on communication as shown in Figure \ref{fig:commTime_lqcd_stencil5d}(a).
Under PAR, LQCD's communication time surges from 17.10ms to 25.51ms, a 49.14\% increase.
In contrast, Stencil5D with a larger peak ingress volume is barely affected by LQCD with less than 3\% communication time variation under all the studied routing algorithms as shown in Figure \ref{fig:commTime_lqcd_stencil5d}(b).
Q-adaptive outperforms adaptive routing algorithms by providing the smallest communication time for both applications in all scenarios.
LQCD only suffers a 9.27\% communication time increase from 13.43ms to 14.67ms.
As a result, Q-adaptive provides better performance for both applications and reduces interference at the same time.

\begin{figure*}[h!]
  \centering
  \includegraphics[scale=\figscale]{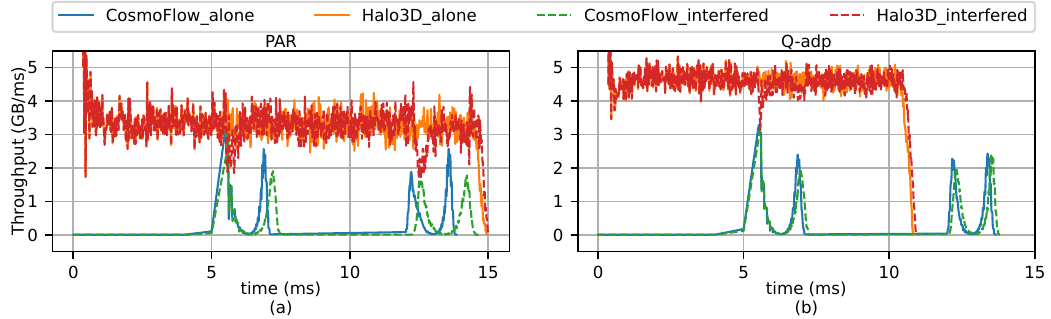}
  \caption{CosmoFlow and Halo3D network throughput along simulated time. The peak of CosmoFlow throughput temporally drags down Halo3D's throughput, but the over interference effect is negligible.}
  \label{fig:random_cosmoflow_halo_bw}
\end{figure*}

\subsection{Impact of Computation} \label{sec:exp_cosmoflow}

\begin{tcolorbox}
Application with long compute time and short communication time can mask communication interference.
\end{tcolorbox}

In Figure \ref{fig:random_app}(d), CosmoFlow exhibits a stable performance between standalone and interfered cases with less communication time variation.
% Compared with Halo3D, background applications with lower injection rate and smaller peak ingress volume 
% % small injection rate and peak ingress volume background applications, 
% such as UR, LU and FFT3D introduce negligible network interference with less than 1.5\% communication time variation.
Compared with Halo3D, background applications such as UR, LU and FFT3D have lower injection rate and smaller peak ingress volume.
These background applications introduce negligible network interference on CosmoFlow with less than 1.5\% communication time increase under all the studied routing algorithms.
A larger interference effect on CosmoFlow is observed when co-running with DL, which has a similar peak ingress volume but a higher injection rate.
CosmoFlow has to spend up to 6.27\% more time on communication with UGALg, but only 1.10\% with Q-adaptive routing.
Halo3D as the highest injection rate application causes 15.93\%, 22.03\%, 21.88\% communication time variation under UGALg, UGALn, and PAR respectively.
However, Q-adaptive can still handle this intensive case very well with 2.90ms communication time, or a 4.88\% variation. 
Figure \ref{fig:random_cosmoflow_halo_bw} depicts CosmoFlow's network throughput under the interference of Halo3D.
Unlike the other applications, CosmoFlow is special because of its long computation interval between communications.
As a result, Halo3D seems to be the only job on the system most of the time with a similar average throughput between the standalone and interfered cases.
In Figure \ref{fig:random_cosmoflow_halo_bw}(b), CosmoFlow reaches its peak throughput of 3.15GB/ms when the first Allreduce operation is issued at 5.51ms.
This high throughput pulse temporally brings down Halo3D's throughput to 2.95GB/ms, which then quickly recovers to its average of 4.51GB/ms.
However, the intermittent decrease in throughput does not affect the overall communication time.

%\subsection{Summary of Application Analysis}

% \textbf{this paragraph should be cut off, or integrated into the key insight boxes} To summarize the pairwise analysis, we observe that when applications are paired together, applications with higher message injection rate tend to be less affected by workload interference.
% Additionally, peak ingress volume is an important metric to capture communication intensity.
% Among the studied workloads, Stencil applications have larger peak ingress volume that delays other applications.
% However, between stencil applications, a higher dimensional stencil patterns with larger peak ingress volume can tolerate communication interference.
% Although communication interference is inevitable due to network sharing, Q-adaptive outperforms adaptive routing in terms of shorter communication time and less performance variation.

\begin{figure*}[ht]
  \centering
    \includegraphics[scale=\figscale]{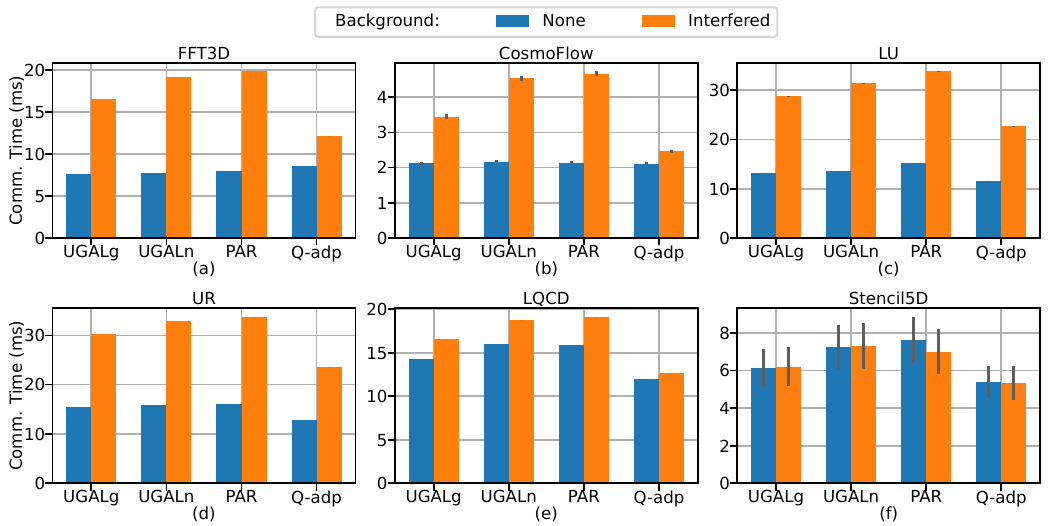}
  \caption{
  Application communication time comparison: ``none" denoting there is no other co-running application and ``interfered" denoting multiple applications in the mixed workloads are co-running in the system.
  }
  \label{fig:random_fullsys_commtime}
\end{figure*}

\section{Mixed workload Analysis} \label{sec:mixedworkload}

In this set of experiments, we go beyond pairwise analysis by co-running multiple applications of different communication patterns, denoted as \emph{mixed workload analysis}. The goal is to analyze workload interference under a mixture of applications. Our analysis focuses on capturing application communication performance under workload interference as well as system-wide network congestion and/or hot spots. 
Table \ref{tab:sys_apps} summarizes the mixed workload,
in which each application takes around one-sixth of the system nodes. LQCD and Stencil5D occupy more compute nodes to construct the high-dimensional process grids. 

\begin{table}[ht]
\centering
\caption{Applications in Mixed Workload Analysis}
\label{tab:sys_apps}
%\begin{tabular*}{\linewidth}{r|rrrrr}
\begin{tabular}{r|rrrrrr}
\hline
Application & FFT3D & CosmoFlow & LU  & UR  & LQCD & Stencil5D \\
\hline
Job size    & 140   & 138       & 140 & 139 & 256 & 243  \\
\hline
\end{tabular}
\end{table}

\subsection{Application Performance Under Workload Interference}

\begin{tcolorbox}
Compared with adaptive routing, Q-adaptive routing reduces workload interference by 49.23\% on average.
Applications with large peak ingress volume can resist interference from other workloads.
\end{tcolorbox}

Figure \ref{fig:random_fullsys_commtime} depicts the network interference from the mixed workload on each individual application.
Among the mixed workload applications, Stencil5D has the largest peak ingress volume and resists network interference with less than 2\% communication delay caused by the other applications.
The second largest peak ingress volume application, LQCD, also has a relative stable performance with an average of 17.87\% communication delay under adaptive routing and only a 6.50\% delay under Q-adaptive routing. 
On average, FFT3D, CosmoFlow, LU, UR and LQCD spend 96.01\% more time on communication under adaptive routing. 
However, these applications experience less communication delay under Q-adaptive routing. 
Specifically, compared with adaptive routing, Q-adaptive results in an average of 49.23\% interference reduction for these applications.
% However, Q-adaptive routing makes these five applications only experience an average of 48.74\% communication delay, which is a 49.23\% network interference reduction compared with adaptive routing solutions.

\subsection{System-wide Network Analysis}

\begin{tcolorbox}
Q-adaptive routing achieves better system-wide performance by providing more balanced traffic distribution, less congestion, and less hot spots.
\end{tcolorbox}

Adaptive routing algorithms are sensitive to local congestion, whereas Q-adaptive routing sends packets based on the overall network condition.
As a result, in the case of a fully loaded system, packets are more likely to be unnecessarily non-minimally forwarded by adaptive routing methods.
This inefficiency consumes more network resources to deliver the same amount of traffic and leads to unbalanced network usage with hot spots. 

\begin{figure}[ht]
  \centering
  \includegraphics[scale=\figscale]{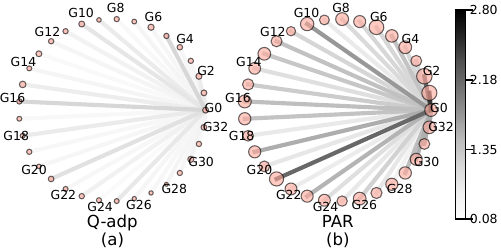}
  \caption{
  Network stall time analysis. Circles denote local links within a group, and a larger circle means higher stall time. 
  Edges between groups denote global links, and a darker color means longer stall time on the global link. 
  For visibility purpose, only global links from Group 0 are plotted. 
  These plots clearly indicate that Q-adpative routing outperforms PAR in terms of reducing network stall time on both local and global links.
  }
  \label{fig:random_fullsys_stalltime}
\end{figure}

Figure \ref{fig:random_fullsys_stalltime} depicts \emph{network stall time} caused by congestion.
Compared with PAR, Q-adaptive forwards packets more efficiently with less congestion shown as the smaller average stall time both within groups (31.42ms vs. 59.15ms) and between groups (0.52ms vs. 1.33ms).
Q-adaptive routing avoids network congestion by having a more balanced system-wide traffic distribution than PAR.
In Figure \ref{fig:random_fullsys_stalltime}(b),
obvious hot spots are observed at Group 1, 2, 10 and 21.
These hot spots also congest the traffic from Group 0 causing a longer global link stall time.
Moreover, among the overly loaded global links, PAR extremely underutilizes G0-G28 global link with only 0.59ms stall time compared with the average of 1.33ms.

\begin{figure}[ht]
  \centering
  \includegraphics[scale=\figscale]{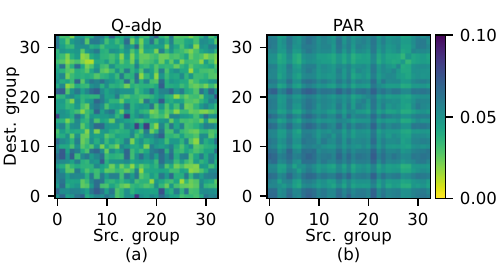}
  \caption{
  Heat map of global link congestion index (non-diagonal dots)  and local link congestion index (diagonal dots).
  PAR results in unbalanced traffic distribution on global and local links with clear diagonal and vertical/horizontal lines. }
  \label{fig:random_fullsys_congestion_index}
\end{figure}

We further analyze system-wide traffic distribution by adapting a metric called \textit{congestion index} from \cite{congestion_index}. It is defined as the ratio between average link throughput and its maximum capacity.
As shown in Figure \ref{fig:random_fullsys_congestion_index}, compared with Q-adaptive routing, PAR generally has a darker color, indicating overall system congestion.
The clear diagonal line in Figure \ref{fig:random_fullsys_congestion_index}(b) demonstrates the unbalanced utilization between local and global links.
Additionally, the pattern of darker horizontal and vertical lines at group 8, 14, 16, 21 and lighter lines at group 2,3,5,6,27,28 unveils the unbalanced global traffic distribution.

\begin{figure}[ht]
  \centering
  \includegraphics[scale=\figscale]{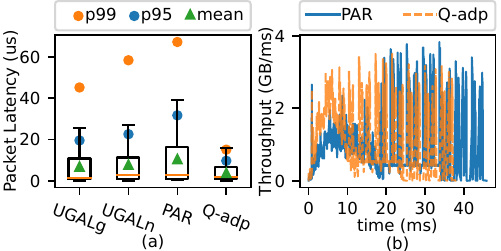}
  \caption{System-wide packet latency distribution and aggregated network throughput under the mixed workload.
  Q-adaptive routing achieves significantly smaller tail latency with higher throughput.}
  \label{fig:random_fullsys_metrics}
\end{figure}

Figure \ref{fig:random_fullsys_metrics} depicts \emph{system-wide packet latency distribution} and \emph{the aggregated network throughput}.
In Figure \ref{fig:random_fullsys_metrics}(a), Q-adaptive has the most concentrated packet latency distribution with the smallest average and tail latency.
Q-adaptive's \SI{3.87}{\micro\second} average packet latency and \SI{15.13}{\micro\second} 99th percentile latency are more than 63\% smaller than those of PAR.
Being able to deliver packets more quickly also leads to higher system throughput.
As shown in Figure \ref{fig:random_fullsys_metrics}(b), Q-adaptive achieves an average throughput of 1.27GB/ms, which is 35.11\% higher than PAR's 0.94GB/ms average throughput.

To conclude, the mixed workload analyses demonstrate that Q-adaptive routing can effectively balance system traffic to avoid network hot spots and congestion.
As a result, when Q-adaptive routing is used, the system can deliver packets more quickly, and therefore yields a higher network throughput, which in turn leads to better application performance. 

\section{conclusion} \label{sec:conclusion}

High-radix, low-diameter Dragonfly interconnect topology is a crucial component for exascale computing.  
A major problem on Dragonfly system is the network competition among co-existing applications for shared network resources (aka workload interference) such that the applications may experience a huge communication delay.
In this study, through pairwise workload analysis and mixed workload network interference study, we have shown that without using any additional complicated techniques, intelligent routing such as Q-adaptive routing can greatly reduce network congestion and save communication time by up to 42.63\%.
We have also presented two metrics to formally quantify an application's communication intensity: message injection rate and peak ingress volume. Several key insights have been presented in our pairwise and mixed workload analyses. 

\section*{Acknowledgment}
We thank the reviewers for their valuable feedback and insightful comments.
This work is supported in part by US National Science Foundation grants CNS-1717763, CCF-2109316, and CCF-2119294.

\bibliographystyle{IEEEtran}
\bibliography{IEEEabrv,reference}

\end{document}